\documentclass[12pt]{iopart}

\usepackage{mathtext}
\usepackage{amsfonts}
\usepackage{amssymb}

\usepackage{bm}
\usepackage{epsfig}
\usepackage{iopams}

\begin{document}

\title{Time modulation of atom sources}

\author{A. del Campo$^1$, J. G. Muga$^1$ and M. Moshinsky$^2$}
\eads{\mailto{qfbdeeca@ehu.es}, \mailto{jg.muga@ehu.es}, \mailto{moshi@fisica.unam.mx}}
\address{$^1$ Departamento de Qu\'imica-F\'isica,
Universidad del Pa\'is Vasco, Apdo. 644, Bilbao, Spain}
\address{$^2$ Instituto de F\'isica, Universidad Nacional Aut\'onoma de M\'exico, 
Apartado Postal 20-364, 01000 M\'exico D.F., M\'exico}
\def\la{\langle} 
\def\ra{\rangle}
\def\om{\omega}
\def\Om{\Omega}
\def\vep{\varepsilon}
\def\wh{\widehat}
\def\tr{\rm{Tr}}
\def\da{\dagger}
\newcommand{\beq}{\begin{equation}}
\newcommand{\eeq}{\end{equation}}
\newcommand{\beqa}{\begin{eqnarray}}
\newcommand{\eeqa}{\end{eqnarray}}
\newcommand{\bV}{{\bf V}}
\newcommand{\bK}{{\bf K}}
\newcommand{\bG}{{\bf G}}

\begin{abstract}
Sudden turn-on of a matter-wave source leads to characteristic oscillations 
of the density profile which are the hallmark feature of diffraction in time. 
The apodization of matter waves relies on the use of smooth aperture functions which 
suppress such oscillations.  
The analytical dynamics of non-interacting bosons arising with different aperture functions are discussed systematically 
for switching-on processes, and for single and many-pulse formation procedures.
The possibility and time scale of a revival of the diffraction-in-time pattern is 
also analysed. 
Similar modulations in time of the pulsed output coupling in atom-lasers    
are responsible for the dynamical evolution and characteristics 
of the beam profile. 
For multiple pulses, different phase schemes and regimes are described and compared. 
Strongly overlapping pulses lead to a saturated, constant beam profile 
in time and space, up to the revival phenomenon.   
\end{abstract}

\pacs{03.75.Pp, 03.75.-b, 03.75.Be}
Coherent matter-wave pulse formation and dynamics have been studied traditionally
in the context of scattering and interferometric experiments with mechanical shutters. 
Suddenly switching-on and off matter-wave sources leads to an oscillatory pattern 
in the particle density which was discovered by one of the authors 
in 1952 and dubbed diffraction in time \cite{Moshinsky52,Moshinsky76}
because of the analogy with the diffraction of a 
light beam from a semi-infinite plane.
The (sudden) ``Moshinsky shutter'' describes the evolution of a truncated 
plane wave suddenly released and admits an analytical solution 
which remains a basic reference for analysing more complex and realistic cases \cite{Kleber94,GV01,GM05,DM05b,DDGCMR06}. 
A wide class of experiments have reported diffraction in time in neutron 
\cite{HFGGGW98} and atom optics \cite{atomdit}, and with electrons 
as well \cite{Lindner05}. 
It was also observed in a Bose-Einstein 
condensate bouncing off from a vibrating mirror, proving that the effect survives even in the presence of a mean-field interaction \cite{CMPL05}. 
A more recent motivation for studying matter-wave pulses 
is the  development of atom lasers: intense, coherent and directed matter-wave 
beams extracted from a Bose-Einstein condensate.   
One of the prototypes uses two-photon Raman excitation   
to create ``output coupled'' atomic pulses with well defined momentum  
which overlap and form a quasi-continuous beam \cite{Phillips99,Phillips00} (For other 
approaches see e.g. \cite{Ketterle97} or \cite{CMRVPG06}).
Clearly, the features of the beam profile are of outmost importance for applications 
\cite{KBMHE05,RGLCFJBA06,KP06}, and will 
depend on the pulse shape, duration, emission frequency, and initial phase.    
While many studies on matter-wave pulse formation and dynamics
deal with single or double pulses, more research is needed to understand and control 
multiple overlapping pulses relevant for current atom-laser experiments.
We shall undertake this objective
within the analytical framework of  
the elementary Moshinsky shutter, here at the level of the 
Schr\"odinger equation for independent bosonic atoms. Non-linear effects will be numerically considered elsewhere. 
This formulation may indeed be adapted also to  
other important features of pulse formation in atom lasers, 
which are different from the mechanical shutter scenario. 
Instead of sudden switching we shall consider smooth aperture functions, 
i.e., ``apodization'', a technique well-known in Fourier optics to avoid the diffraction effect
at the expense of broadening the energy distribution \cite{Fowles68}.
Manifold applications of this technique have also been found in time filters for signal analysis \cite{BT59}.  

In more detail, motivated by the recent 
realisation of an atom laser in a waveguide \cite{Guerin06}, 
we shall focus on the longitudinal beam features and explore the 
space-time evolution of effective one dimensional sources represented 
by ``source'' boundary conditions of the form
\beq\label{source0} 
\psi(x=0,t)=\chi(t)\,e^{-i\om_0 t}, \qquad \forall t>0.
\eeq
The aperture function $\chi(t)$ modulates the wave amplitude 
at the source and is therefore responsible for the apodization in time, 
that is, the suppresion of the fringes in the beam profile.
However we shall show that, for long enough pulses or for multiple 
overlapping pulses,
such effect holds only in a given time domain 
because of the revival of the diffraction in time.   

Localised sources \cite{BT98,MB00} represented by an amplitude $\psi(x=0,t)\sim e^{-i\om_0 t}$
have been used for understanding tunnelling 
dynamics or front propagation and related time quantities such as the tunnelling times \cite{MSE02}. 
Moreover, this approach has shed  new light on transient 
effects in neutron optics \cite{GG84,FGG88,HFGGGW98} and diffraction of atoms both in 
time and space domains \cite{BZ97,DM05}. 
The connection and equivalence between ``source'' boundary conditions, in which the state is specified at a single point at all times and the usual initial value problems, in which the state is specified  for all points 
at a single time, was studied in \cite{BEM01}. 
Notice that we shall use the source conditions (\ref{source0}) for simplicity, 
but the pulses could also be examined with the initial value approach.

In section \ref{turning-on} we shall review the sudden 
switching on of the source since any other case is a combination of this 
elementary dynamics. Single pulse formation 
with standard apodizing functions is examined in sec. \ref{onepulse}, 
and we describe the dynamics for a smoothly switched-on source
in sec. \ref{switching}; Section \ref{mp} is devoted to the multiple pulse case  
treated with different aperture functions, and the paper ends with a discussion.

\section{Turning-on the source\label{turning-on}}

We shall consider the dynamics of a source of frequency $\om_0=\hbar k_0^2/2m$, which is turned on in a horizontal waveguide. This has the advantage of cancelling the effect of gravity off, 
avoiding the decrement of the de Broglie wavelength $\lambda_{dB}=2\pi/k_0$ due to the downward acceleration \cite{Guerin06}. Note that the effects of an external linear potential on the diffraction in time phenomenon can also be taken into account following \cite{DM06a}.
At the same time the dynamics is effectively one-dimensional provided that $\om_0\ll\om_{\perp}$ where $\om_{\perp}$ is the transverse frequency of the waveguide.

The sudden approximation for turning on a source is well-known \cite{Moshinsky52,MB00,BEM01,MSE02} but briefly reviewed here for completeness. 
The aperture function is then taken to be 
the Heaviside step function, $\chi_0(t)=\Theta(t)$.
The inverse Fourier transform of the source 
$\psi_0(x=0,k_0,t)=e^{-i\om_0 t}\Theta(t)$ is given by
\beqa
\widehat{\psi}(x=0,k_0,\omega)&=&F^{-1}[\psi_0](x=0,k_0,\om)
\equiv\frac{1}{\sqrt{2\pi}}\int_{-\infty}^{\infty}\psi_0(x=0,k_0,t)e^{i\om t}dt\nonumber\\
&=&\frac{i}{\sqrt{2\pi }}\, 
\frac{1}{\omega-\omega_{0}+i0}.
\eeqa
For $x,t>0$, the wave function evolves according to 
\begin{displaymath}
\psi_0(x,k_0,t)=\frac{i}{2\pi }\int_{-\infty}^{\infty}
d\omega\,\frac{e^{i k x-i\omega t}}{\omega-\omega_{0}+i0},
\end{displaymath}
where ${\rm{Im}} k\ge 0$.   
This can be written in the complex $k$-plane by deforming the contour of
integration to $\Gamma_{+}$ which goes from 
$-\infty$ to $\infty$ passing above the poles, 
\begin{eqnarray}
\psi_0(x,k_0,t)&=&\frac{i}{2\pi }
\int_{\Gamma_+}dk\,2 k\,\frac{e^{i k x-i
\frac{\hbar k^{2}t}{2 m}}}{k^{2}-k_{0}^{2}}\nonumber\\
&=&
\frac{i}{2\pi }\int_{\Gamma_+}\!\!\!dk\, 
\left(\frac{1}{k+k_{0}}\!+\!\frac{1}{k-k_{0}}\right)
e^{i k x-i \frac{\hbar k^{2}t}{2 m}}.
\nonumber
\end{eqnarray}
Since one of the definitions of the Moshinsky function $M(x,k,t)$ 
is precisely
\beq
M(x,k,t)=\frac{i}{2\pi }\int_{\Gamma_+}dk'
\frac{e^{i k' x-i \frac{k'^{2}t}{2}}}{k'-k},
\eeq
(The Moshinsky function can be related 
to the complementary error function, see \ref{AppA}.), 
the resulting state can then be simply written as
%
\beq
\label{sudden}
\psi_{0}\left(x,k_0,t\right)=M\left(x,k_0,\frac{\hbar t}{m}\right)
+M\left(x,-k_0,\frac{\hbar t}{m}\right).
\eeq
Such combination of Moshinsky functions is ubiquitous when working with 
quantum sources and will appear throughout the paper.
The source density profile corresponding to Eq.(\ref{sudden}) exhibits the 
characteristic oscillations of diffraction in time \cite{Moshinsky52, Moshinsky76}.
We are interested in describing more general switching procedures. However, given 
their dependence on pulse formation results, its discussion will be 
postponed to section \ref{switching}.

\section{Single pulse}\label{onepulse}
In this section we study the formation of a single-pulse of duration $\tau$ from a quantum source 
modulated according to a given aperture function $\chi^{(1)}(t)$, the superscript 
denoting the creation of a single pulse.
Notice that some apodizing functions were already explored 
in the quantum domain \cite{DM05}, but we shall present a more general description.

In particular, let us consider the family of aperture functions 
$\chi_{n}^{(1)}(t)=\sin^{n}(\Om t)\Theta(t)\Theta(\tau-t)$ with $\Om=\pi/\tau$, see Fig. 1(a).
The inverse Fourier transform of the source amplitude at the origin is given by
\beqa
\widehat{\psi}_n^{(1)}(x=0,k_0,\om)=\frac{1}{\sqrt{2\pi}}\frac{2^{-n}e^{i(\om-\om_0)\tau/2}\tau\Gamma(1+n)}
{\Gamma\Bigg[\frac{(2+n)\pi+(\om-\om_0)\tau}{2\pi}\Bigg]
\Gamma\Bigg[\frac{(2+n)\pi-(\om-\om_0)\tau}{2\pi}\Bigg]},
\eeqa 
where $\Gamma(z)$ is the Gamma function \cite{AS65}.
The energy distribution of the associated wavefunction \cite{BEM01}
is proportional to $\om^{1/2}|\widehat{\psi}_n^{(1)}(x=0,\om)|^2$. 
As shown in Fig.1(b), the smoother the aperture function the wider is 
the energy distribution, which will affect the spacetime profile of the resulting pulse.
We shall next illustrate the dynamics for the $n=0,1,2$ cases. The solution for an arbitrary 
aperture function is presented in \ref{AppB}.
\subsection{Rectangular aperture function}
For the case $\chi_{0}^{(1)}(t)=\Theta(t)\Theta(\tau-t)$  corresponding to a single-slit in time
the dynamics of the wave function is given by 
\beqa
\psi_{0}^{(1)}(x,k_0,t;\tau)=\psi_0\left(x,k_0,t\right)-
\Theta(t-\tau)e^{-i\om_0 \tau}\psi_0\left(x,k_0,t-\tau\right).
\eeqa
Note that if $t<\tau$, before the pulse has been fully formed, the problem reduces to that discussed 
in the previous section, being $\chi(t)=\Theta(t)$. Equation (\ref{pulse})  
is also valid for these times,  
in contrast to previous works restricted to $t>\tau$ \cite{GG84,FGG88,BZ97,DM05}.
The same will be true for the following pulses.
\subsection{Sine aperture function}
Let us consider now $\chi_{1}^{(1)}(t)=\sin(\Omega t)\,\Theta(t)\,\Theta(\tau-t)$. 
Then, defining $\om_{\pm}=\om_{0}\pm\Om$ and the corresponding wavenumbers 
$k_{\pm}=\sqrt{2m\om_{\pm}/\hbar}$, 
\beqa
\psi_1^{(1)}(x,k_0,t;\tau)=\frac{i}{2}\sum_{\alpha=\pm}\alpha\psi_{0}^{(1)}(x,k_{\alpha},t;\tau).
\eeqa
\subsection{Sine-square aperture function} 
The sine-square (Hanning) aperture function is given by
\beqa\label{Hanning}
\chi_2^{(1)}(t)&=&\sin^{2}({\Om t})\,\Theta(t)\,\Theta(\tau-t)
=\frac{1}{2}\left[1-\frac{1}{2}\cos({2\Omega t})\right]\,\Theta(t)\,\Theta(\tau-t).
\eeqa
Introducing $k_{\beta}=\sqrt{2m(\om_{0}\pm2\Om)/\hbar}$ 
with $\beta=\pm$, leads to 
\beqa
\psi_2^{(1)}(x,k_0,t;\tau)=\frac{1}{2}\left[\psi_{0}^{(1)}(x,k_0,t;\tau)
-\frac{1}{2}\sum_{\beta=\pm}\psi_{0}^{(1)}(x,k_{\beta},t;\tau)\right],
\eeqa
where the effect of the apodization is to substract to the pulse
with the source momentum
two other matter-wave trains associated with $k_{\beta}$, all of 
them formed with rectangular aperture functions.

\begin{figure}
\begin{center} 
\includegraphics[height=5cm,angle=-90]{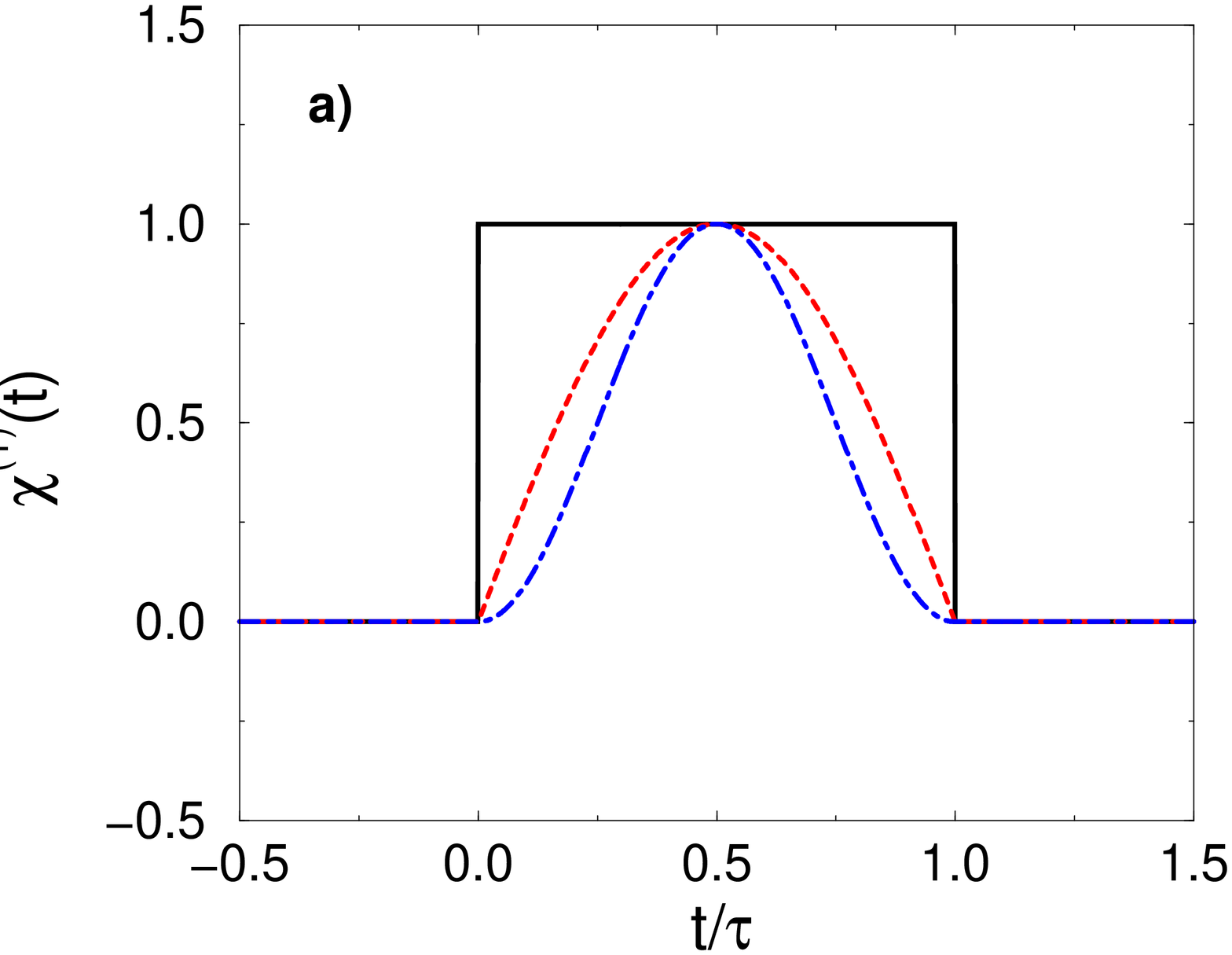}
\includegraphics[height=5cm,angle=-90]{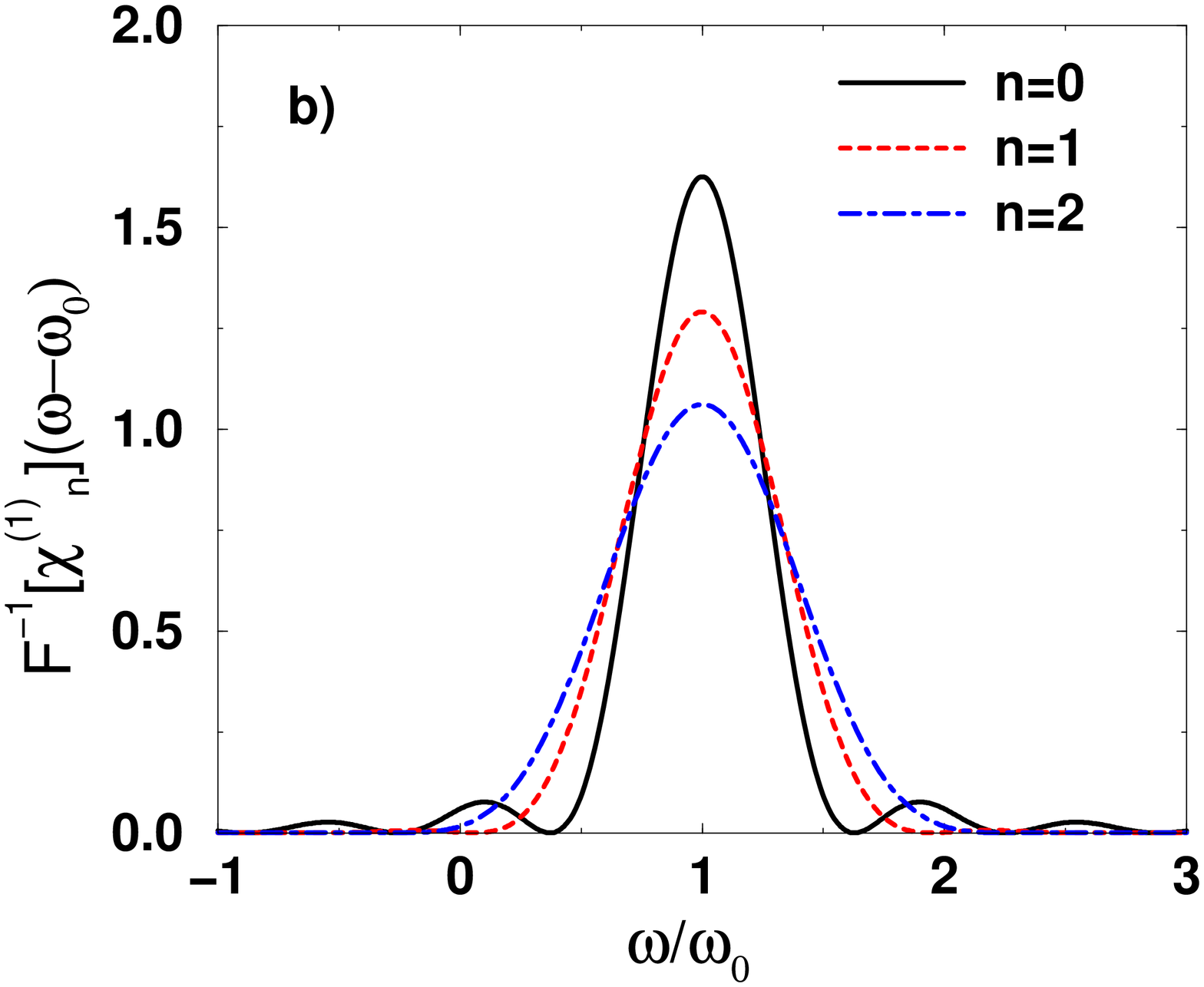}
\vspace*{.2cm}
\caption{\label{apop}
a) Family $\chi_n(t)$ of single-pulse aperture functions. 
The smoothness at the edges 
increases with in this order: rectangular ($n=0$, continuous line), 
sine ($n=1$, dashed line), sine-square ($n=2$, dot-dashed line). 
b) The aperture function in the frequency domain, $F^{-1}[\chi_n^{(1)}](\om-\om_0)$,
broadens as a result of apodization, with increasing $n$.
}
\end{center}
\end{figure}

%
%
\begin{figure}
\begin{center} 
\includegraphics[height=4cm,angle=-0]{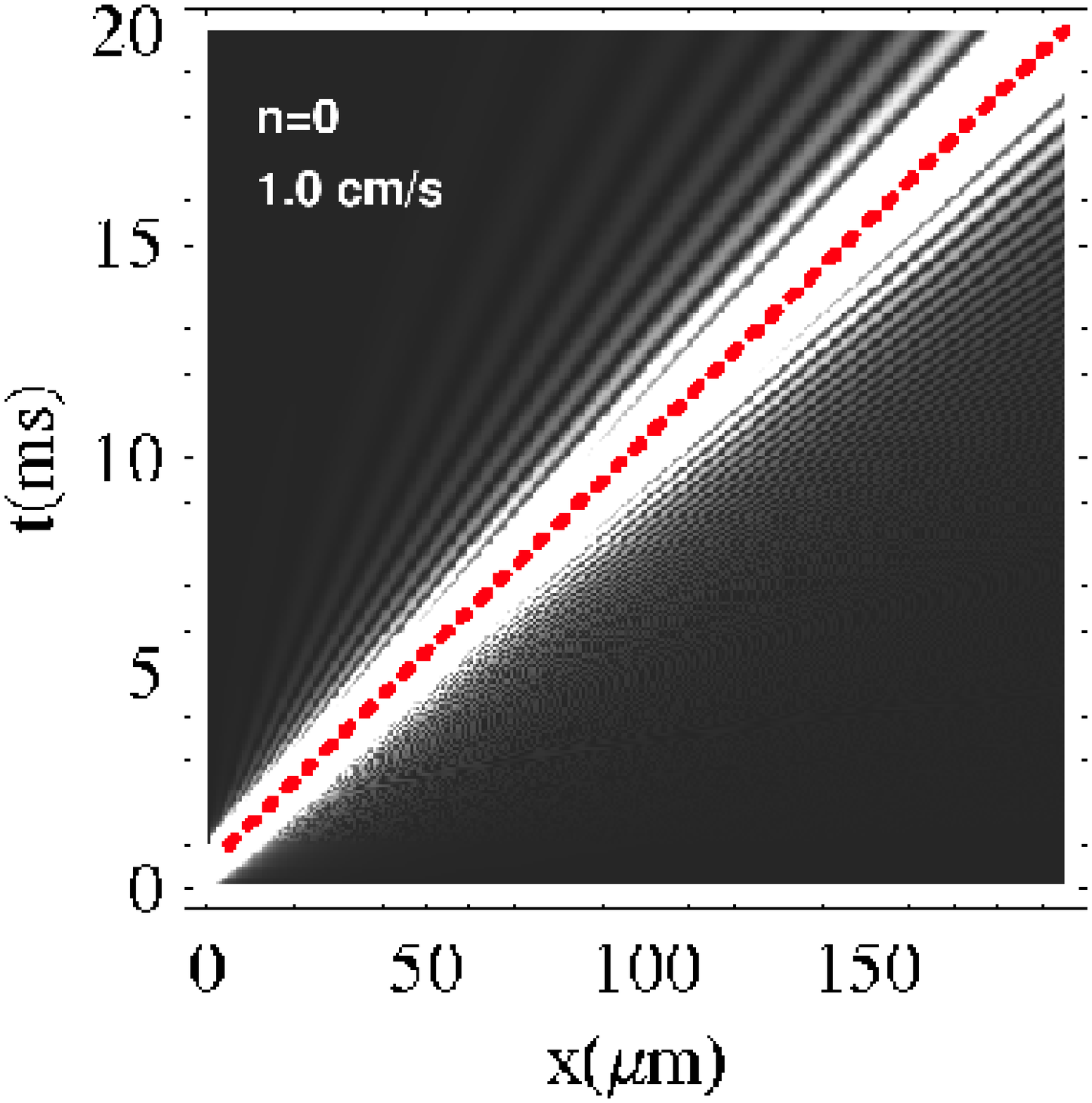}
\includegraphics[height=4cm,angle=-0]{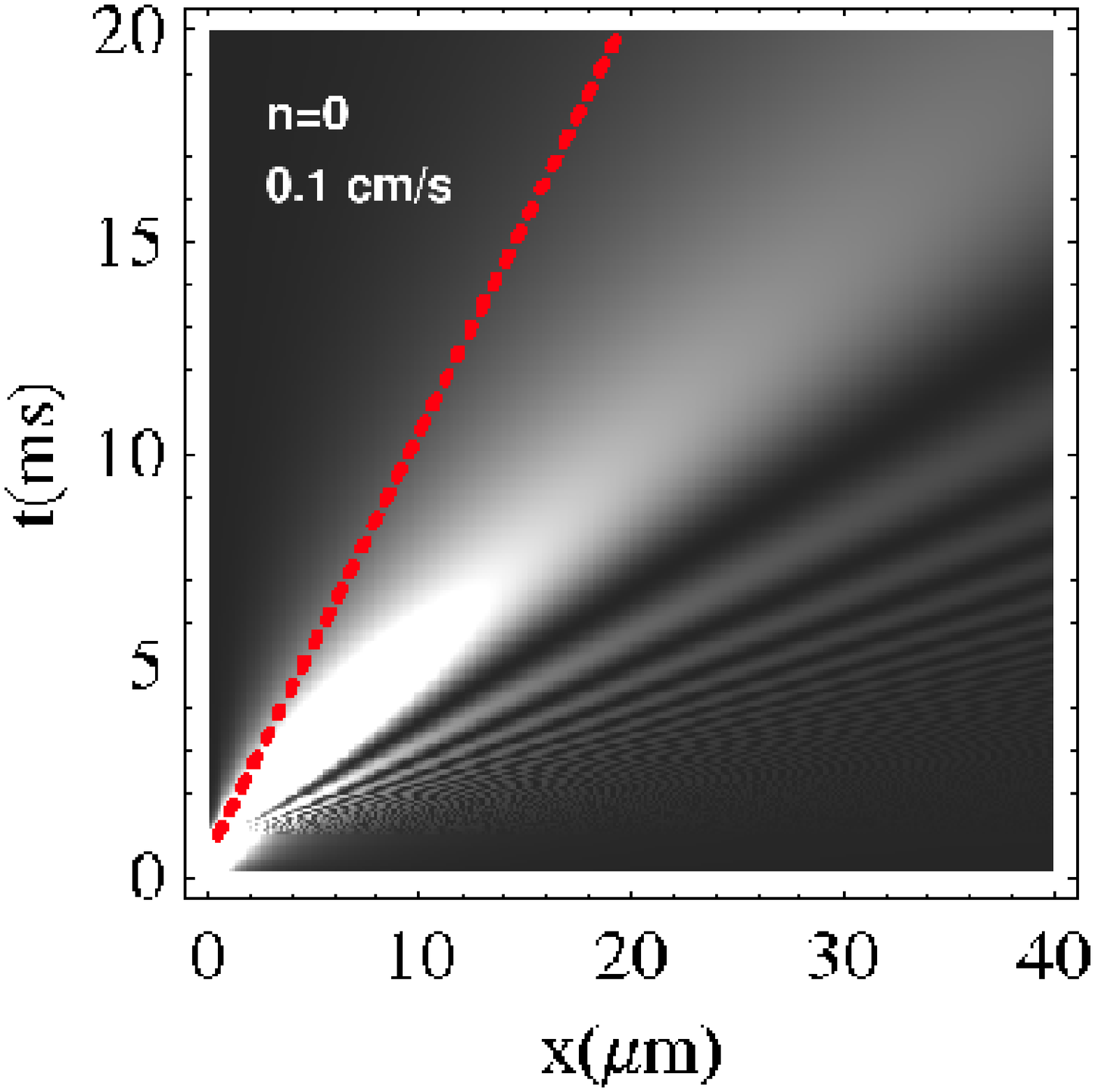}

\includegraphics[height=4cm,angle=-0]{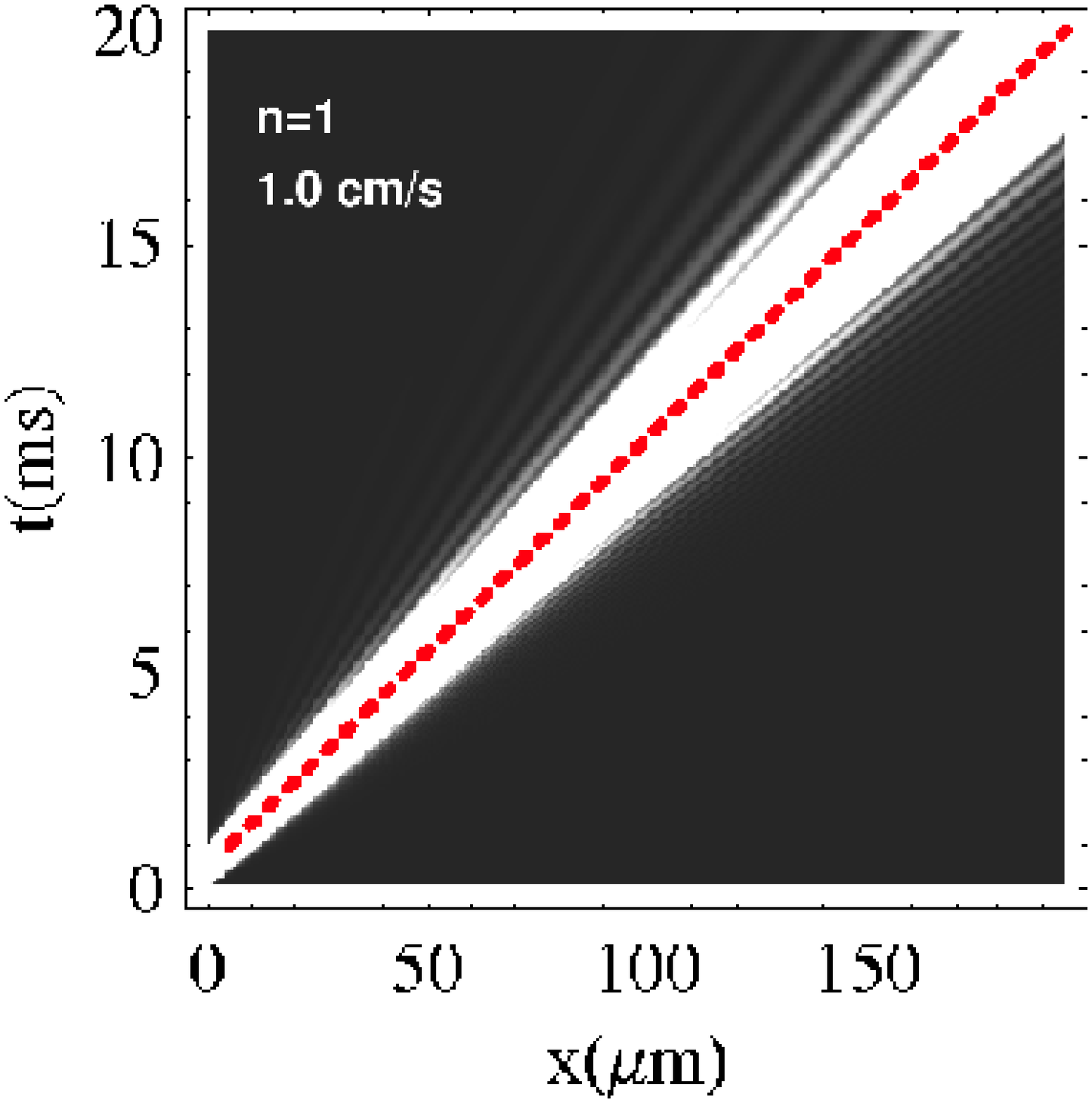}
\includegraphics[height=4cm,angle=-0]{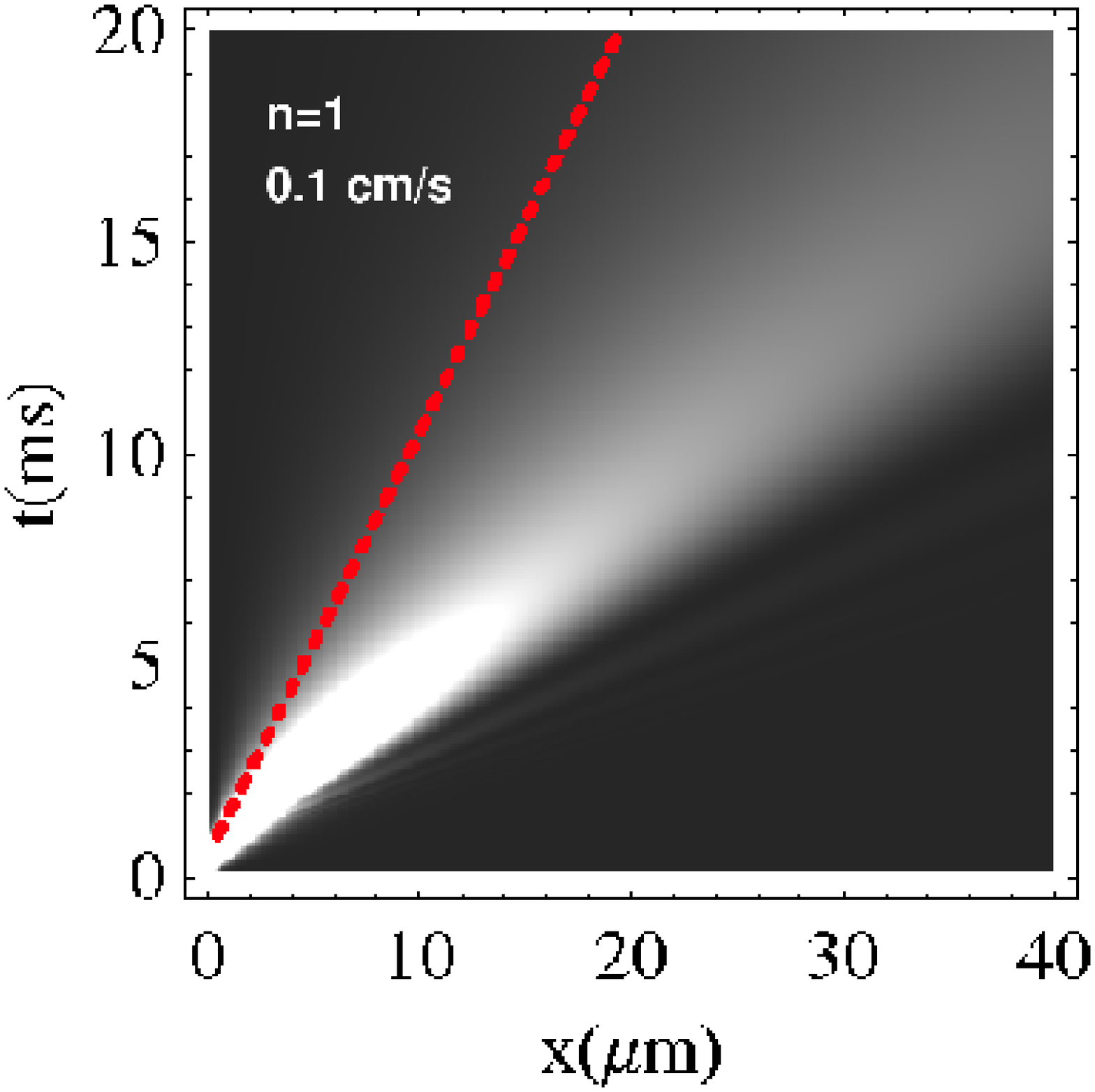}

\includegraphics[height=4cm,angle=-0]{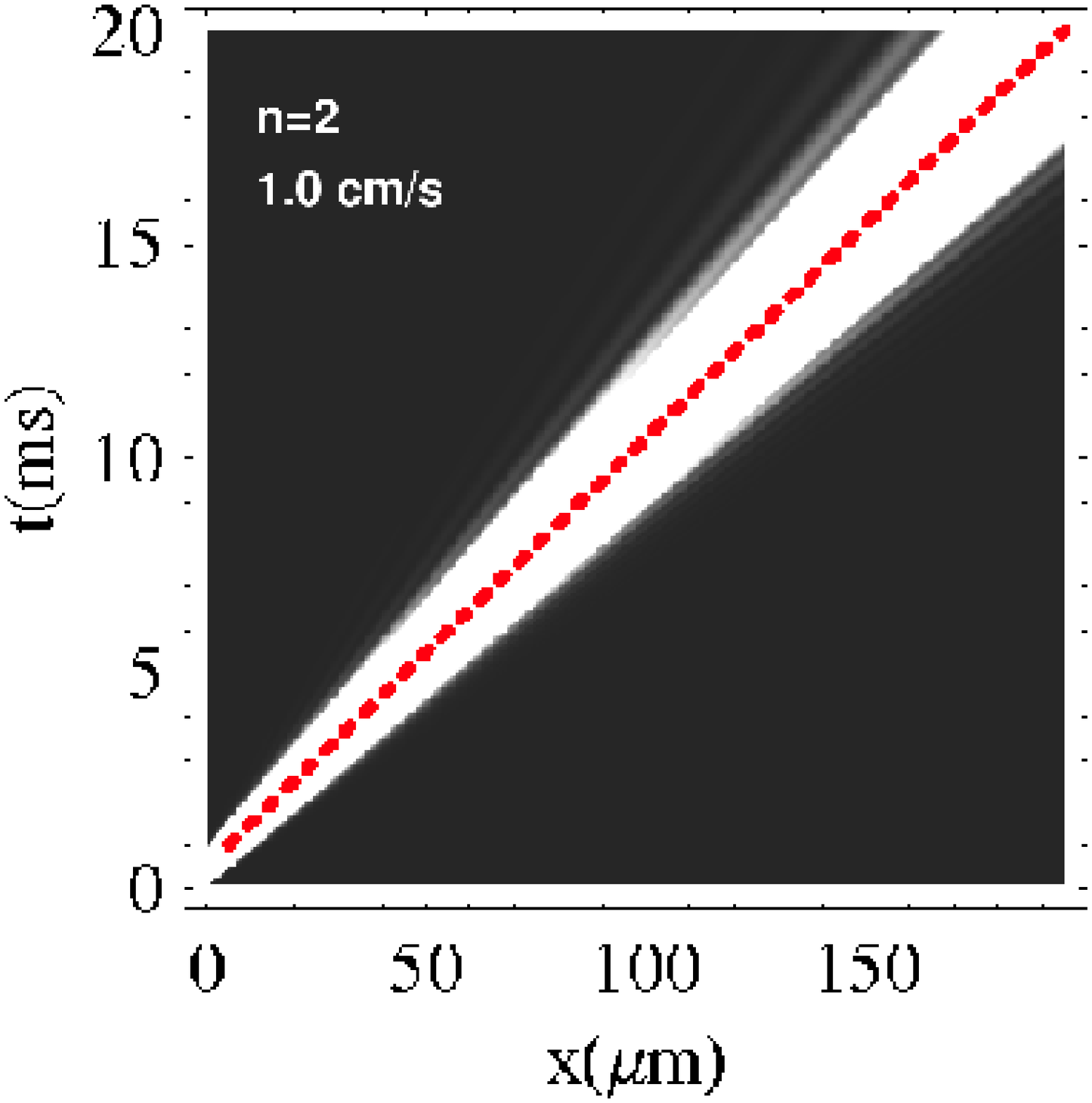}
\includegraphics[height=4cm,angle=-0]{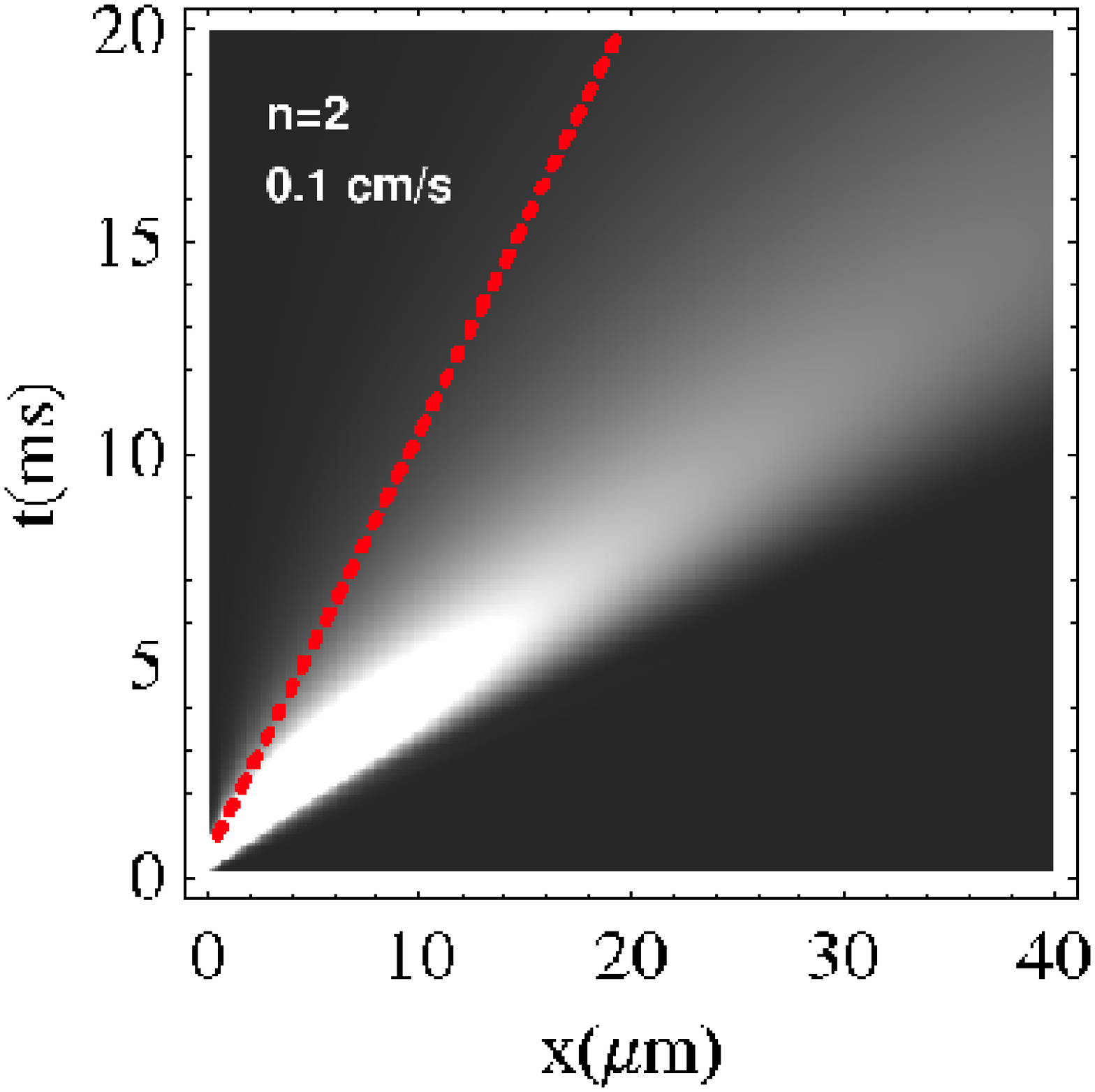}
\vspace*{.2cm}
\caption{\label{xpulses}
(color online) Spacetime effects of the time-energy uncertainty relation. 
A pulse of duration $\tau<\top_0$ is sped up with respect to the source frequency. 
The spacetime dynamics is illustrated for a $^{87}$Rb source with $\tau=1$ms with 
$\hbar k_0/m=1$cm/s (left column, $\top_0=0.0467$ms) 
and $\hbar k_0/m=0.1$cm/s (right column, $\top_0=4.67$ms), 
 modulated with different apodizing functions to form a single pulse.
The dashed line reproduces the classical trajectory. The grey scale 
changes from dark to light as the function values increase.
Figures in the same column show the suppression of the sidelobes in 
the probability density with increasing $n$.
}
\end{center}
\end{figure}

\begin{figure}
\begin{center} 
\includegraphics[height=8cm,angle=-90]{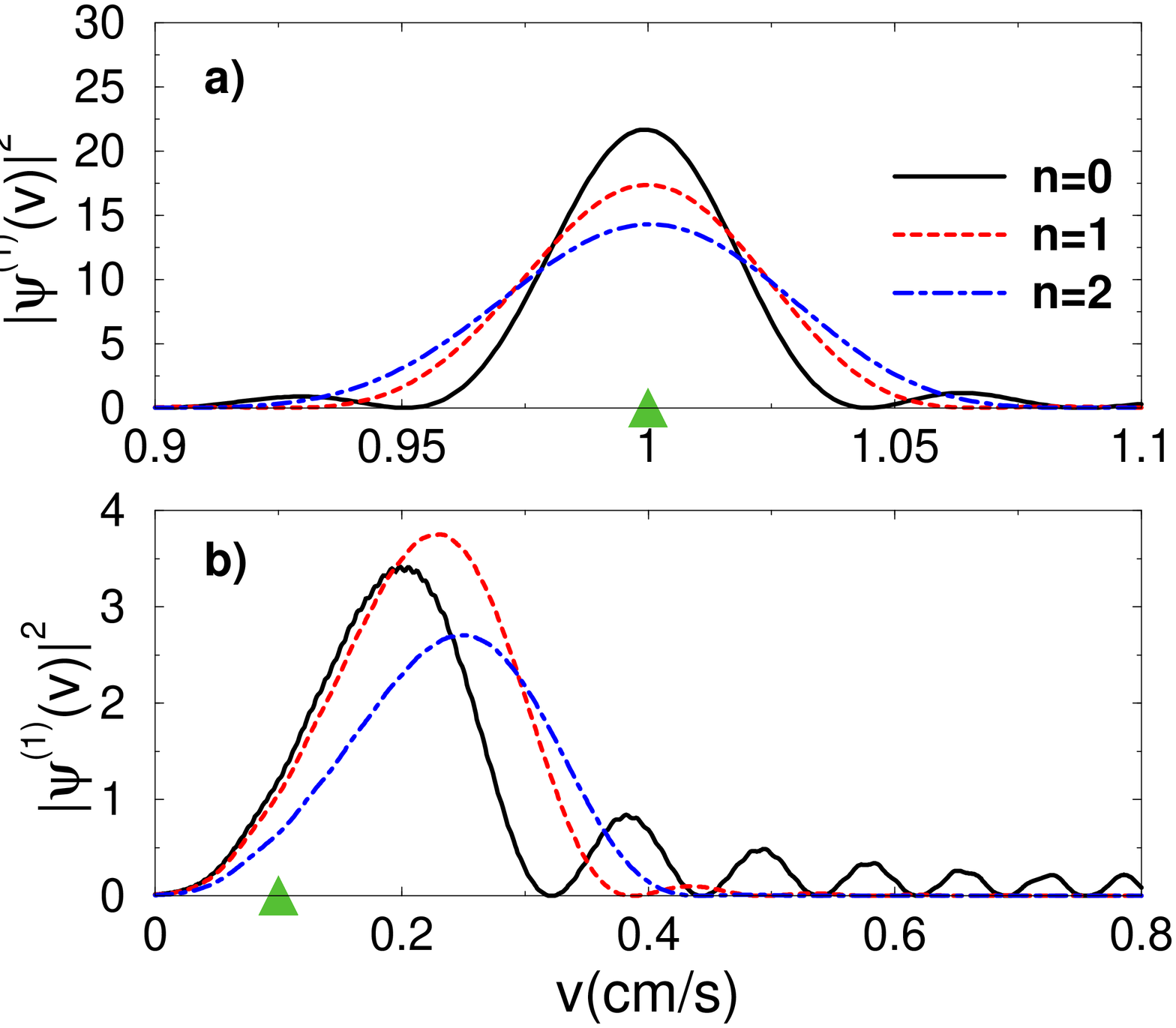}
\vspace*{.2cm}
\caption{\label{vpulses}
(Color on-line) Distortion of the velocity distribution for pulse duration smaller than $\top_0$.
The corresponding pulses to Fig. \ref{xpulses} are plotted in velocity space 
for a $^{87}$Rb source with $\tau=1$ms apodized with different aperture functions and
a) $\hbar k_0/m=1$cm/s ($\top_0=0.0467$ms) and b) $0.1$cm/s ($\top_0=4.67$ms).
The triangle marks the classical velocity. 
}
\end{center}
\end{figure}

The effect of a short $\tau$ on the energy distribution has already been studied in 
\cite{Moshinsky76, DM05} from which a time-energy uncertainty relation was inferred. 
Indeed, the uncertainty product reaches an approximate  minimum of $2\hbar$ if 
the pulse duration is shorter than $\top_0=\pi/\om_{0}$.
The space-time dynamics for a source apodized with a rectangular aperture function 
is illustrated in Fig. \ref{xpulses}, 
where it is shown that for $\tau>\top_0$ the quantum average of the position operator 
follows the classical free-particle trajectory whereas if $\tau\lesssim\top_0$ the quantum 
pulse is sped up. 
Fig. \ref{vpulses} shows for different $\chi_n(t)$ ($n=0,1,2$) the effect on the 
velocity distribution which is only centered at $\hbar k_0/m$ for $\tau>\top_0$.
Concerning the apodization, the higher the value of $n$ the narrower is 
the mean width of $\chi_n(t)$ and the larger the shift on the mean velocity. 
The smoothing of the $\chi_n(t)$ is reflected on the suppression of the sidelobes 
in the velocity distribution. 
\section{Smooth switching: apodization and diffraction in time}\label{switching}
\begin{figure} 
\begin{center}
\includegraphics[height=6cm,angle=-90]{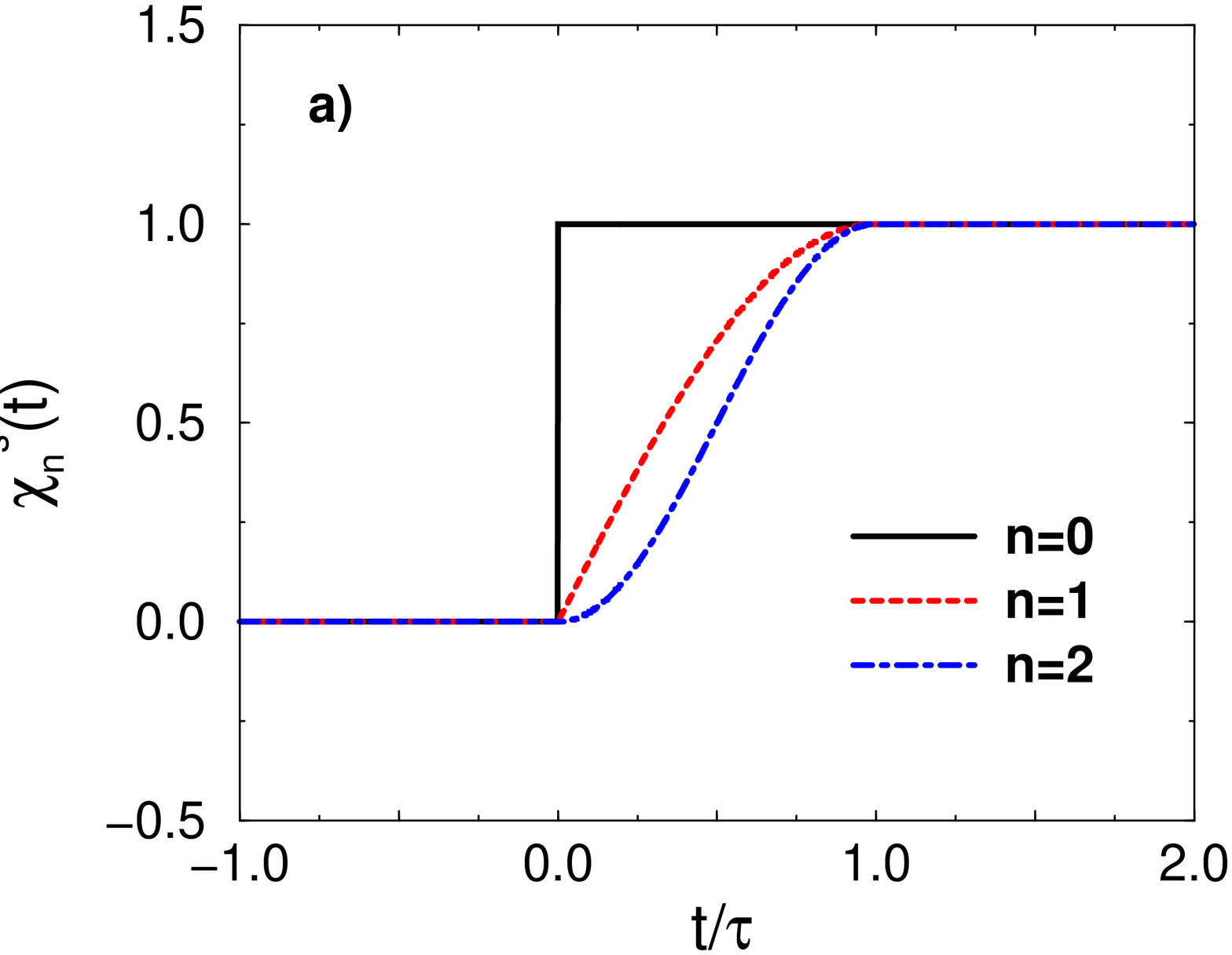}
\includegraphics[height=5.7cm,angle=-90]{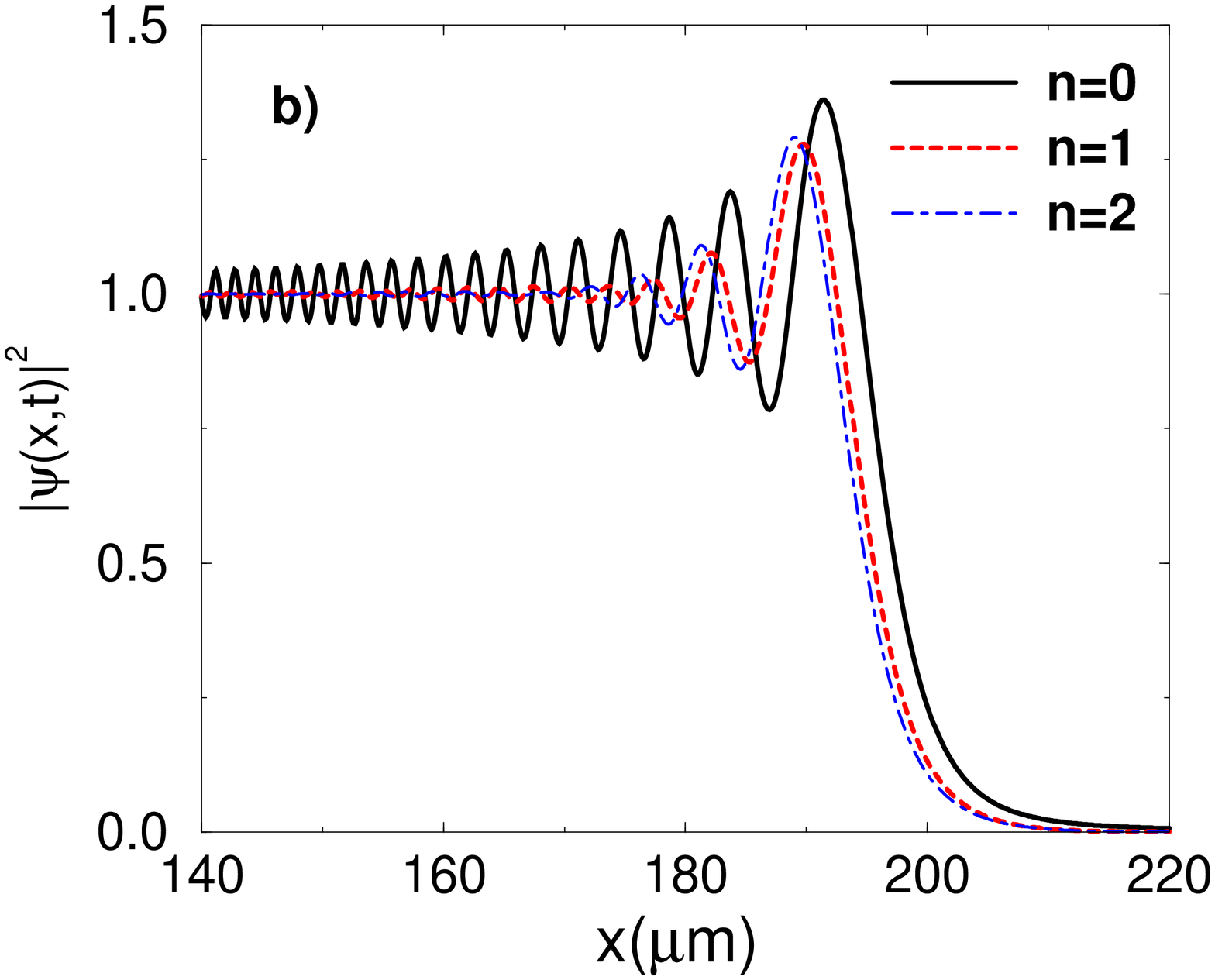}
\vspace*{.2cm}
\caption{\label{switchfunc}
a) Family $\chi_n(t)$ of switching-on aperture functions.  
The smoothness at the edges 
increases with $n$.
b) Apodization of the $^{87}$Rb beam profiles with increasing smoothness of the aperture function (higher $n$), 
and fixed switching time $\tau_s=0.5$ms at $t=20$ms, and $\hbar k_0/m=1$ cm/s.
}
\end{center}
\end{figure}

Instead of forming a finite pulse as discussed in the previous section,  
some sources are prepared to reach a stationary regime with constant flux 
after the initial transient. The extreme case is the sudden, zero-time 
switching of Sec. \ref{turning-on}.  
In this section we shall study how the matter-wave dynamics 
is modified for slow aperture functions with a given switching time $\tau$. 
In order to do so, we introduce the family of switching functions
\begin{displaymath}
\label{switch}
\chi^{s}(t)=
\cases{
 0, \qquad t<0\\
\chi(t), \qquad 0\leq t<\tau\\
1, \qquad t\geq\tau}
\end{displaymath}
where it is assumed that $\chi(\tau)=1$, and which is represented in Fig. (\ref{switchfunc})a 
for $\chi(t)=\sin^n\Om_s t$ with $n=0,1,2$ and $\Om_s=\pi/2\tau$.
Note that for $n=0$, one recovers the sudden aperture $\chi_0(t)=\Theta(t)$ which 
maximises the diffraction in time fringes.
For $n=1$ (n=2) the source amplitude follows half a sine (square)-lobe.
The general result for  the time-dependent wavefunction simply reads,
\beqa
\label{spsi}
\psi_\chi^s(x,k_0,t;\tau)
=e^{-i \om_0 \tau}\psi_0\left(x,k_0,t-\tau\right)\Theta(t-\tau)
+\psi^{(1)}_{\chi}(x,k_0,t;\tau),
\eeqa
where the superscript $s$ stands for ${\it switching}$ and $\psi^{(1)}_{\chi}$ 
refers to a pulse of length $\tau$ and apodizing function $\chi$.

%
%
%
\begin{figure} 
\begin{center}
\includegraphics[height=6cm,angle=-90]{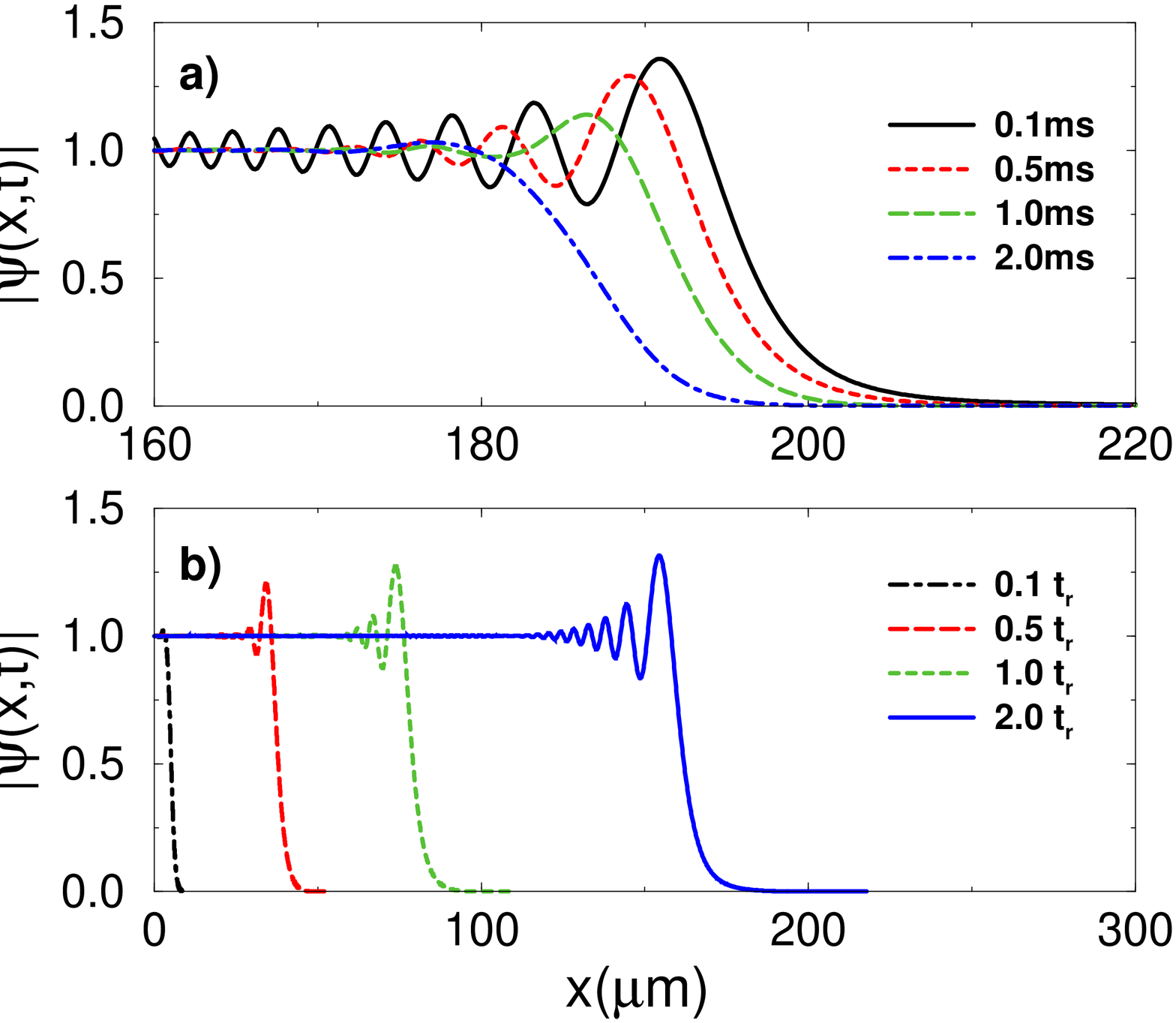}
\caption{\label{apofunc}
Revival of the diffraction in time.
a) The oscillations in the beam profile characteristic of the 
diffraction in time are gradually suppressed with increasing switching time $\tau$.
The beam profile is shown at $t=20$ms, and $\hbar k_0/m=1$cm/s. For increasing $\tau$ the amplitude of 
the oscillations diminishes and the signal is delayed.
b) During the time evolution there is a revival of the diffraction 
in time which is again clearly visible for times greater than 
$t_r=\om_0\tau^2$ ($\tau=1$ms, $\hbar k_0/m=0.5$cm/s, and $t_r=16.8$ms).
Both a) and b) refer to a $^{87}$Rb source which is switched on following a sine-square function.
	
}
\end{center}
\end{figure}
%

\begin{figure} 
\begin{center}
\includegraphics[height=6cm,angle=-90]{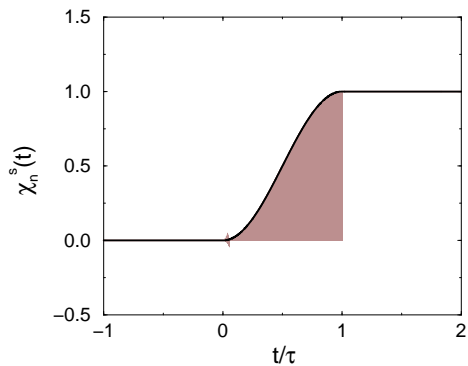}
\caption{\label{cap}
Aperture function decomposition.
In a switching proccess the aperture function has a finite transient component 
responsible for the apodization (shaded area). Such apodization 
cap is followed by a step excitation for $t\geq\tau$ which leads eventually to the revival of diffraction in time. 
}
\end{center}
\end{figure}

%
Figure \ref{switchfunc}(b) shows
the effect of a finite switching time on 
the oscillatory pattern at a given time.
As a result of the smooth aperture functions new frequencies are singled out, 
leading to a progressive suppression of diffraction in time. 
Moreover, for a given $\chi_n(t)$ function, a larger switching time $\tau$ 
increases the apodization of the source, 
washing out the fringes in the probability density, see Fig. \ref{apofunc}(a). 
(In this sense, the effect of a finite band-width source 
is tantamount to a time dependent modulation, 
as was shown in \cite{MB00}.)
However, it is remarkable that the effect of the apodization is
limited in time 
because of a revival of the diffraction in time. 
The intuitive explanation is that the intensity of the signal from the apodization ``cap'', 
see Fig. \ref{cap}, decays with time whereas the intensity of the main 
signal (coming from the step excitation) remains constant. 
For sufficiently large times, the main signal, carrying its 
diffraction-in-time phenomenon, overwhelms the effect of the 
small cap. 
One can estimate the revival time by considering that the 
momentum width of the apodization 
cap will be of the order   
\beq
\Delta p\approx
\sqrt{2m\hbar(\om_0+\Om_s)}-\sqrt{2m\hbar(\om_0-\Om_s)}
\approx\sqrt{2m\hbar\om_0}\frac{\pi}{2\om_0\tau},
\eeq
and taking into account the, essentially linear with time , spread of the 
cap half-pulse described by $\psi_{\chi}^{(1)}$ in Eq. (\ref{spsi}) (we have checked numerically 
that the classical dispersion relation $\Delta x(t)\simeq\Delta x(0)+\Delta pt_r/m$ holds). 
%
%

The ratio of the spatial widths corresponding to the initial and evolved cap $\psi_{\chi}^{(1)}$ is given by
\beq
R=\frac{\Delta x(0)}{\Delta x(t_r)}\approx\frac{\Delta x(0)}{\Delta x(0)+\Delta pt_r/m}.
\eeq

Taking $\Delta x(0)=\sqrt{2\hbar\om_0/m}\tau$ and imposing $R=1/2$ leads to 
the expression for a revival time scale 
\beq
t_r\approx \om_0\tau^2,
\eeq
which is analogous to the Rayleigh distance of classical diffraction theory 
but in the time domain \cite{Brooker03}.
The smoothing effect of the apodizing cap 
cannot hold for times much longer than $t_r$.
This is shown in Fig. \ref{apofunc}(b), where the initially 
apodized beam profile develops eventually spatial fringes, which reach the maximum value associated 
with the sudden switching at $t\approx 2t_r$.
\section{Multiple pulses}\label{mp}
We show next that the single pulse aperture functions $\chi^{(1)}(t)$, 
described in section \ref{onepulse} 
are useful to consider matter wave sources periodically modulated in time.
Indeed, the $N$-pulse aperture function can be written as a convolution of 
$\chi^{(1)}(t)$ with a grating function.
In particular, if the grating function $g(t)$ is chosen to be a finite Dirac comb 
 $g_N(t)=\sum_{j=0}^{N-1}\delta(t-j T)$, it is readily found 
that 
\beq
\chi^{(N)}(t)=\chi^{(1)}(t)*g_N(t)=\int dt'\chi^{(1)}(t-t')g_N(t')=\sum_{j=0}^{N-1}\chi^{(1)}(t-jT),
\eeq
which describes the formation of $N$ consecutive $\chi^{(1)}$-pulses, 
each of them of duration $\tau$ and with an ``emission rate'' (number of pulses per unit time) $1/T$.
%
%

If we are to describe an atom beam periodically modulated with such  apodizing function, 
the space-time evolution simply reads, 
\beq
\label{first}
\psi_{\chi}^{(N)}(x,k_0,t)=\sum_{j=0}^{N-1}e^{-i\om_0 jT}\psi_{\chi}^{(1)}(x,k_0,t-jT;\tau)
\Theta(t-jT)
\eeq
being a linear combination of the single-pulse wavefunction 
$\psi^{(1)}(x,t)$ studied in section \ref{onepulse}.  
The $\Theta(t-jT)$ corresponds to the condition that the $j$-th 
pulse starts to emerge only after $jT$.


A similar mathematical framework can be employed to describe an atom laser
in the no-interaction limit.
The experimental prescription for an atom laser depends on the outcoupling mechanism.
Here we shall focus on the scheme described in \cite{Phillips99,Phillips00}.
The pulses are obtained from a condensate trapped in a MOT, and a well-defined momentum is 
imparted on each of them at the instant of their creation through a
stimulated Raman process.
The result is that each of the pulses  $\psi_{\chi}^{(1)}$ does not have a memory phase, 
the wavefunction describing the coherent atom laser being then  
\beq
\phi_{\chi}^{(N)}(x,k_0,t)=\sum_{j=0}^{N-1}\psi_{\chi}^{(1)}(x,k_0,t-jT;\tau)
\Theta(t-jT), 
\eeq
compare with (\ref{first}). 

\begin{figure} 
\begin{center}
\includegraphics[height=6cm,angle=-90]{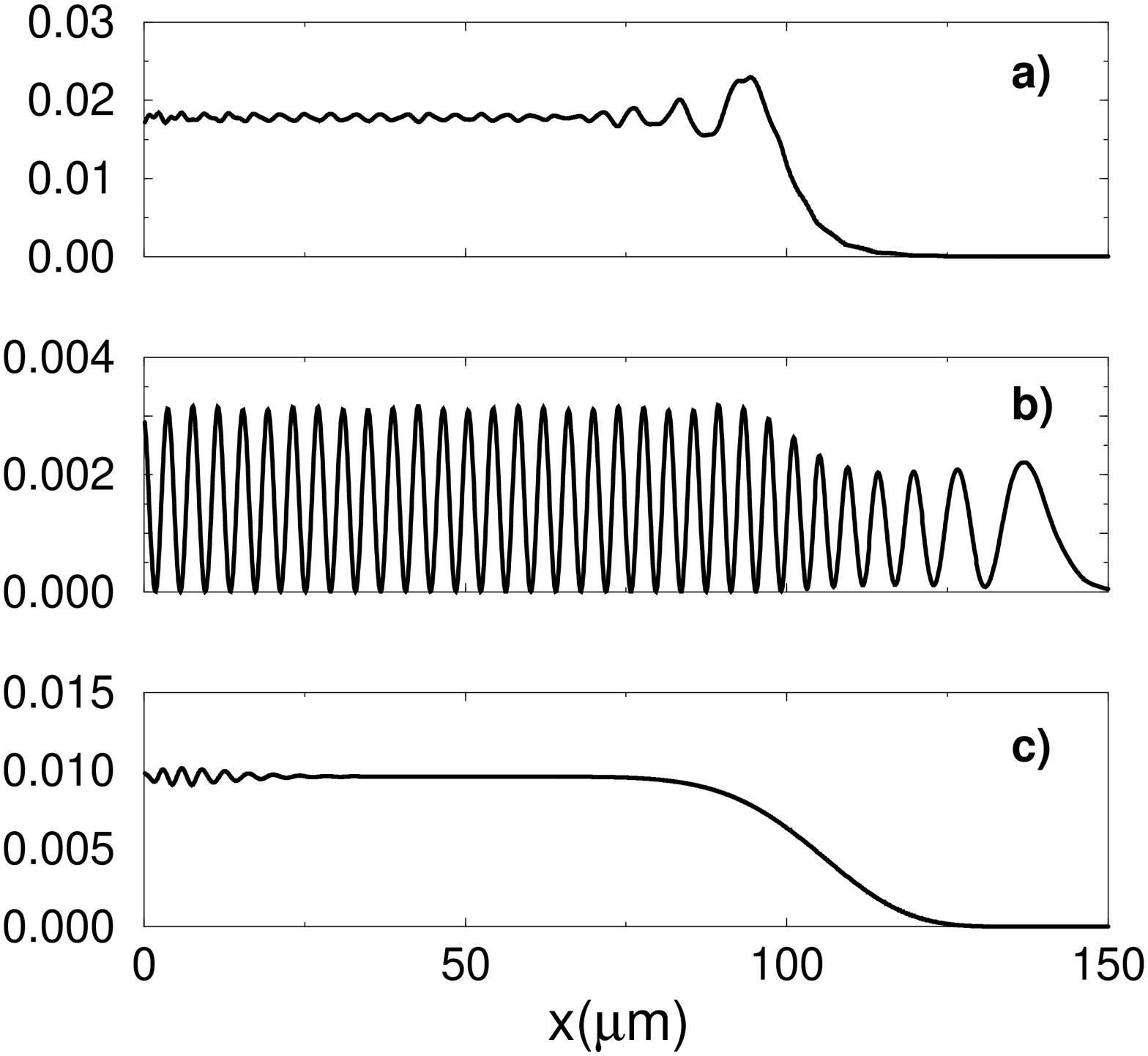}
\includegraphics[width=5.3cm,height=7cm,angle=-90]{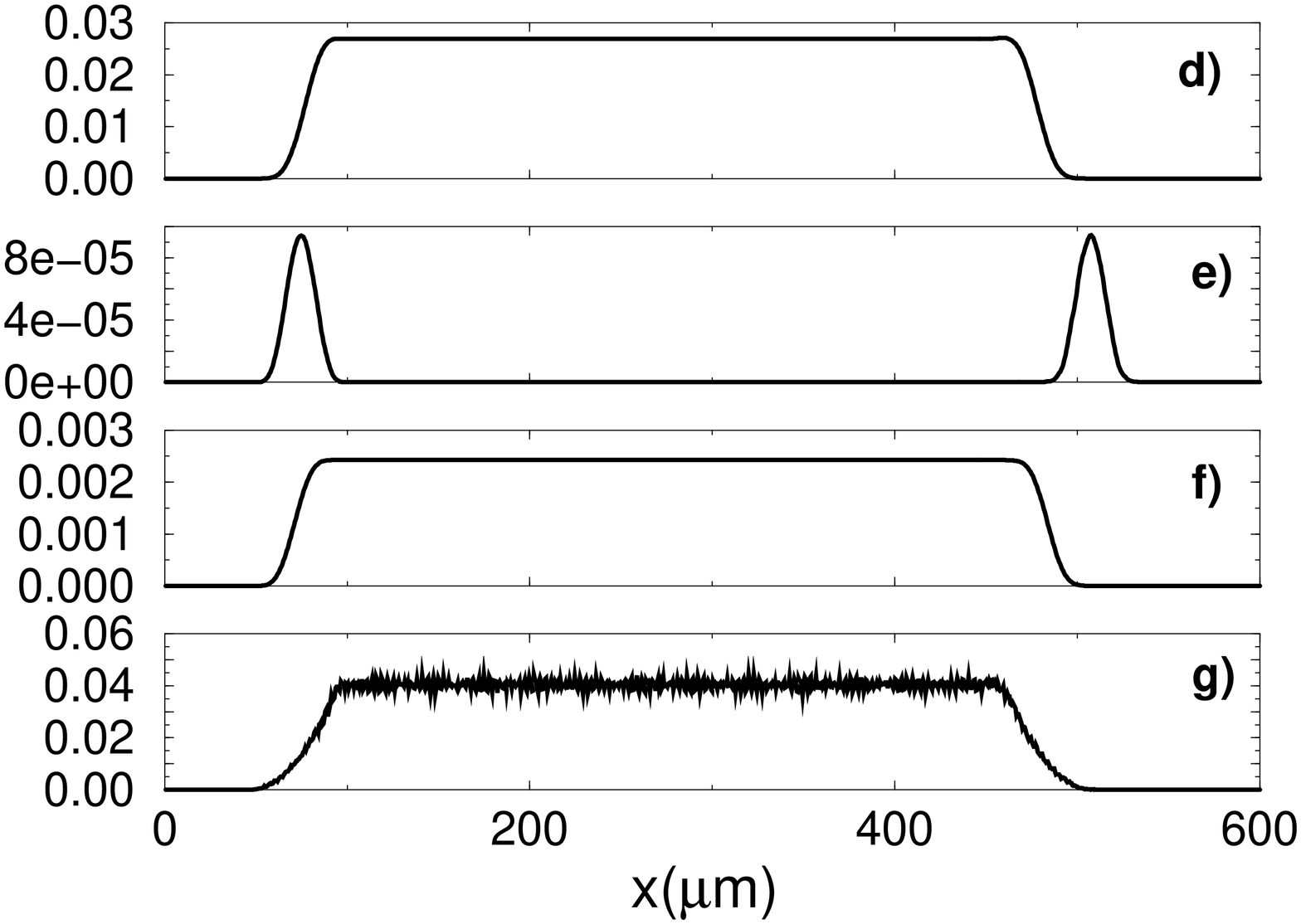}
\hspace*{.2cm}
\caption{\label{atoml}
Beam density profile of a $^{23}$Na atom laser with $\tau=0.833ms$, $T=0.3ms$, in the
 coherently constructive $\om_0T=10\pi$ ($5.9$ cm/s) (a),
 destructive  $\om_0T=11\pi$ ($5.9$ cm/s) (b), and incoherent case (c), 
with same conditions than a)  with a sine-square aperture function.
The conditions equivalent to the experiment reported in \cite{Phillips99} 
($\tau=0.83$ms, $T=0.05$ms) are studied 
for the constructive (d), destructive (e) and incoherent case (f).
If the pulse is not apodized the beam profile becomes noisy as shown in 
(g) for rectangular $\chi$, in the same conditions than (d). 
In all cases the norm is relative to the incoherent case.
}
\end{center}
\end{figure}
It is interesting to impose a relation between the outcoupling period $T$ and 
the kinetic energy imparted to each pulse $\hbar\om_0$, such that $\om_0 T=l\pi$, 
where if $l$ is chosen an even (odd) integer 
the interference in the beam profile $\psi_{\chi}^{(N)}$ 
is constructive (destructive), see Fig. \ref{atoml}.
However,  in the case of a periodically chopped atom beam $\psi_{\chi}^{(N)}$,
Eq. (\ref{first}), 
the additional phase $e^{-i\om_0 jT}$ singles out the constructive interference 
independently of the parity of $l$.

So far we have assumed that the source is coherent. The 
interference pattern between adjacent pulses 
is affected by the presence of noise in the phases \cite{VVDHR06}.
The average over many realisations in which the phase $\alpha_j$ 
of each pulse varies randomly in $[0,2\pi]$ 
leads to the density profile,
\beqa
\label{inco}
\rho(x,t)=\left\la\left\vert\sum_{j=0}^{N-1}e^{i\alpha_j}\psi_{\chi}^{(1)}(x,k_0,t-jT;\tau)
\Theta(t-jT)\right\vert^{2}\right\ra_{\alpha}
\eeqa
which 
reproduces the incoherent sum 
of the pulses, $\rho(x,t)=\sum_{j}\vert\psi_{n}^{(1)}(x,k_0,t-jT;\tau)\vert^{2}\Theta(t-jT)$.

One advantage of employing stimulated Raman pulses is that any desired fraction of atoms 
can be extracted from the condensate. Let us define $s$ as the number of atoms outcoupled in a single pulse, $\psi^{(1)}_{\chi}$. 
The total norm for the $N$-pulse incoherent atom laser is simply $\mathcal{N}_{inc}=sN$.
A remarkable fact for coherent sources is that the total number of atoms $\mathcal{N}_c$
outcoupled from the Bose-Einstein condensate reservoir depends on the nature of the interference \cite{Felipe}.
Therefore, in Figs \ref{atoml} and \ref{atltime} we shall consider the signal relative to the incoherent case, namely, $\vert\psi^{(N)}_{\chi}\vert^{2}/\mathcal{N}_{inc}$.

Note the difference in the time evolution between coherently 
constructive and incoherent beams in Fig. \ref{atltime}: 
%
\begin{figure} 
\begin{center}
\includegraphics[height=8cm,angle=-90]{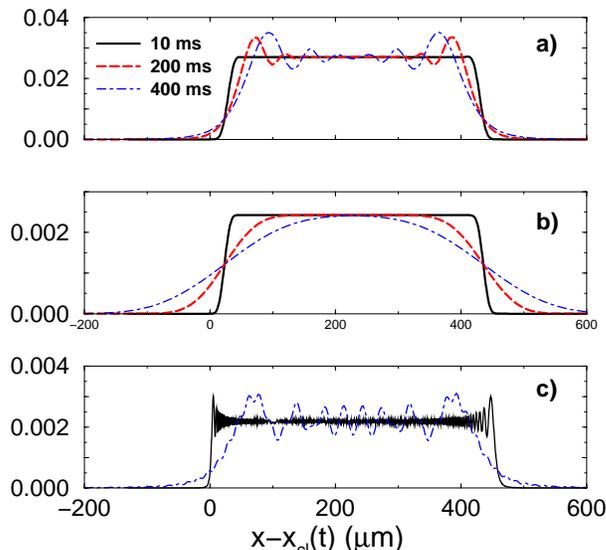}
\hspace*{.2cm}
\caption{
\label{atltime}
Time evolution of a $^{23}$Na atom laser beam modulated by a sine-square function with $\tau=0.833$ms, $T=0.05$ms, moving at 
$5.9$cm/s in the a) coherently constructive, b) incoherent case, and c) rectangular single pulse approximation to a) (the snapshot at $t=200$ms is omitted for clarity). 
Note that $x_{cl}=\hbar k_0 t/m$ is the classical trajectory.
In all cases the norm is relative to the incoherent case, see (\ref{inco}) and the paragraph bellow.
}
\end{center}
\end{figure}
%
At short times both present a desirable saturation in the probability density of the beam.
However, in the former case a revival of the diffraction in time phenomena takes place in a time scale $\om_0\tau^{2}$, similar to the smooth switching considered in section \ref{switching}, 
whereas the later develops a bell-shape profile.
\begin{figure} 
\begin{center}
\includegraphics[height=5cm,angle=-90]{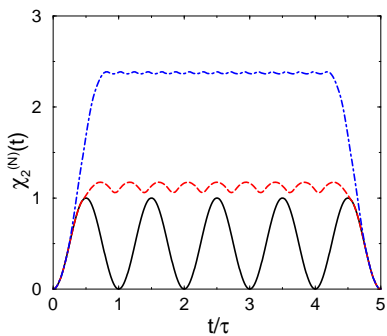}
\hspace*{.2cm}
\caption{
\label{chin}
(color online) Many-pulse aperture function $\chi_2^{(N)}(t)$ with $\tau_{eff}=5\tau$ and increasing number of pulses N: 5, (solid line), 10 (dashed line), and 20 (dot-dashed line). In the highly overlapping regime $\chi_2^{(N)}(t)$ can be approximated by a a single pulse rectangular aperture function of duration $\tau_{eff}=(N-1)T+\tau$.
}
\end{center}
\end{figure}
%

Interestingly, whenever the overlap amongst different pulses 
is coherently constructive or with phase memory (as in the periodic chopping of a beam), 
an effective single-pulse aperture function can be introduced. 
Indeed, if $\tau/T\gg1$ as in the actual experiments \cite{Phillips99,Phillips00}, 
it follows that $\chi_{eff}(t)=\sum_{j=1}^{N-1}\chi(t-jT)\propto\Theta(t)\Theta(\tau_{eff}-t)$ 
for any $\chi$, where the effective pulse duration is $\tau_{eff}=(N-1)T+\tau$. More precisely it reaches a plateau in the time interval $[\tau,nT-\tau]$ but has both a front and a back cap of duration $\tau$ associated with the switching on and off. This is graphically illustrated in Fig. \ref{chin} for $\chi_2^{(N)}(t)$, describing the formation of $N$ pulses with a sine-square modulation $\chi_2^{(1)}(t)$. Moreover, the quality of this approximation improves as the time of evolution becomes greater than the revival time, see Fig \ref{atltime}(c).
Clearly this is only valid in the highly-overlapping regime whereas if the overlap is restricted to few pulses no saturation of the beam profile is observed. 
This is an important simplification to optimise the aperture function 
and design the beam profile of an atom laser.

%
%
%
%

%
%
%
\section{Discussion}
We have studied the dynamics of non-interacting boson sources which are 
modulated in time with arbitrary aperture functions. 
Starting with different switching-on procedures we have accounted 
first for single pulse formation and systematically described the multiple pulse case.
Whenever the modulation is sudden, the spreading of the source exhibits 
oscillation in the density profile, which is the taletelling sign 
of the quantum diffraction in time phenomenon. 
For apodized sources with smooth aperture functions the oscillations 
are suppressed, thus bringing to the quantum domain the classical apodization techniques 
usual in signal analysis and Fourier optics.
However, for long or multiple overlapping pulses 
the smoothing is limited by a time scale $t_r$, associated with a revival 
of the diffraction-in-time phenomenon.
Multiple pulse formation has been examined for different possible cases. 
For constructive or for random averaged phases, a saturation effect occurs 
providing a flat intensity profile before $t_r$, which later  develops 
spatial fringes and a bell-shape profile respectively. 
For both types of beam formation the initial flat beam region is smooth for
apodized pulses but noisy for square ones. 
While these results are already quite relevant for an understanding of pulse formation and 
intended applications of the atom laser, the next natural step will be to extend the present analysis to systems with atom-atom interaction.

\ack{A. C. acknowledges the hospitality of the members of IF-UNAM during the completion of this work. 
The work has been supported by CONACYT (40527F), Ministerio de Educaci\'on y Ciencia
(BFM2003-01003 and FIS2006-10268-C03-01), and UPV-EHU (00039.310-15968/2004).
A.C. acknowledges financial support by the Basque Government (BFI04.479).  
}

\begin{appendix}

\section{The Moshinsky function}\label{AppA}

The Moshinsky function can be related to the Faddeyeva $w(z)$  
and complementary error function \cite{Faddeyeva,AS65},
\beq
M(x,k,\tau):=\frac{e^{i\frac{x^{2}}{2 \tau}}}{2}w(-z),
\qquad {\rm where}\qquad 
z=\frac{1+i}{2}\sqrt{\tau}\left(k-\frac{x}{\tau}\right),
\eeq
and $w(z)$ is defined as
\beq
w(z):= e^{-z^{2}}{\rm{erfc}}(-i z)=\frac{1}{i\pi}\int_{\Gamma_{-}}du
\frac{e^{-u^{2}}}{u-z},
\eeq
$\Gamma_{-}$ being a contour in the complex $z$-plane which goes from
$-\infty$ to $\infty$ passing below the pole.

\section{Arbitrary aperture function}\label{AppB}

In what follows we will be interested in pulses modulated with an 
arbitrary aperture function $\chi(t)$ which is zero outside the interval 
$[0,\tau]$. 
An analytical result can be obtained in such case by means of Fourier series.
%
Noting that
\beq
\chi(t)=\sum_{r=-\infty}^{\infty}c_{r}e^{i\frac{2\pi rt}{\tau}}\Theta(t)\Theta(\tau-t)
\quad {\rm with}\qquad
c_{r}=\frac{1}{\tau}\int_{0}^{\tau}\chi(t)e^{-i\frac{2\pi rt}{\tau}}dt, 
\eeq
the  wavefunction at the origin can be conveniently written for all times 
as an infinite series
\beqa \label{eq:sbc}
\fl\psi_{\chi}(x=0,k_0,t) = \sum_{r=-\infty}^{\infty}c_{r}
e^{-i \left(\omega_{0}+\frac{2\pi rt}{\tau}\right) t}\Theta(t)\Theta(\tau-t)
=\sum_{r=-\infty}^{\infty}c_{r}e^{-i \omega_{r} t}\Theta(t)\Theta(\tau-t),
\eeqa
%
%
%
which evolves for $x,t>0$ as 
\begin{eqnarray}
\psi_{\chi}(x,k_0,t;\tau) 
& = & \sum_{r=-\infty}^{\infty}c_{r}\psi_{0}^{(1)}(x,k_r,t;\tau)
\end{eqnarray}
where 
\beqa\label{pulse}
\psi_{0}^{(1)}(x,k_r,t;\tau)=\psi_0\left(x,k_r,t\right)-
\Theta(t-\tau)e^{-i\om_r \tau}\psi_0\left(x,k_r,t-\tau\right)
\eeqa
and $\hbar k_{r}=\sqrt{2m\hbar\om_r}$, with the branch cut taken along the negative
imaginary axis of $\om_{r}=\om_{0}+2\pi r/T$.
This is an amusing expression for it relates the dynamics of a source with an arbitrary 
aperture function with that of rectangular pulses of different coherent sources. 
As a caveat, less general decompositions may be simpler as in the $\chi_1^{(1)}(t)$ 
case discussed in section \ref{onepulse}.

\end{appendix}

\end{document}